\documentclass[aps,pra,preprint,groupedaddress,showkeys,floatfix]{revtex4}

\usepackage[dvips]{graphicx}
\begin{document}

\begin{center}
{ \Large Phase Effects in Two-Photon Free-Free Transitions in a
Bichromatic Field of Frequencies $\omega$ and $3\omega$}

\vspace*{.5cm}

Aurelia Cionga and Gabriela Zloh

\vspace*{.25cm}

Institute for Space Sciences, P.O. Box MG-23,
Bucharest, R-76900 Romania\\
\end{center}

\vspace*{.5cm} \noindent{\bf Abstract.} The effect of the relative
phase between the components of a bichromatic field of frequencies
$\omega$ and $3\omega $ is discussed in the case of free-free
transitions in laser-assisted electron-hydrogen scattering. For
fast projectile and low field intensities, the role of target
dressing is pointed out.

\vspace*{.5cm}

\section{Introduction}

High harmonic generation techniques have recently been used to
provide intense sources of multichromatic coherent radiation
(L'Huillier et al 1992). Atomic and molecular systems are
submitted to such radiation fields in order to get new insights
into the dynamics of laser assisted processes. Also, they may
represent sensitive "tools" (V{\'{e}}niard et al 1996), which
allow us to  investigate a key parameter in high harmonic
generation phenomenon: the phase difference between the harmonics.

Free-free transition in laser-assisted electron-atom scattering in
a bichromatic field is a process in which the phase difference is
a sensitive parameter. Theoretical investigations on this topic
have recently been published.
 The early results were obtained for low frequencies,
neglecting the dressing of the target (Varr{\'o} and Ehlotzky
1993, 1993a, and Ehlotzky 1994). A major aspect investigated  in
these works was the influence of the relative phase between the
components of the bichromatic field on the laser assisted signals.
However, perturbative calculations for both monochromatic (Dubois
et al 1986, Kracke et al 1994) and bichromatic fields (Cionga \&
Buic\u{a} 1998) have shown that the dressing of the target by the
radiation field plays an important role when the field frequency
is no longer small. The effect of target dressing on free-free
transitions in a bichromatic field of frequencies $\omega $ and
2$\omega $ was investigated in the domain of moderate field
intensities for fast projectiles: the laser-atom interaction was
described by first order perturbation theory (Varr{\'o} and
Ehlotzky 1997) or by second order perturbation theory (Cionga and
Zloh 1999).

The aim of our work is to study two-photon free-free transitions
in electron-hydrogen scattering when the radiation field is the
superposition of two components of frequencies $\omega $ and
3$\omega $:
\begin{equation}
\vec{{\cal A}}(t)=\vec{\varepsilon}{\cal A}_0\cos \omega t+{\vec{\varepsilon}%
}^{\;\prime }{\cal A^{\prime }}_0\cos \left( 3\omega t+\varphi \right) .
\label{c2}
\end{equation}
${\cal A}_0$ is the amplitude of the vector potential describing
the laser field of frequency $\omega $, ${\cal A^{\prime }}_0$
describes the second harmonic of frequency
$\omega^{\prime}=3\omega $; ${\vec{\varepsilon}}$ and
${\vec{\varepsilon}}^{\;\prime }$ are the corresponding
polarization vectors. $\varphi $ is the relative phase between the
harmonic and the fundamental field and we focus our attention on a
systematic investigation of its influence on laser assisted
signals which correspond to the  scattered electrons with the
final energy
\begin{equation}
E_f=E_i\pm 2\omega,  \label{ener}
\end{equation}
where  $E_{i(f)}$ is the initial (final) energy of the projectile.
In the presence of the radiation field (\ref{c2}) the energy
(\ref{ener}) is reached by two different quantum paths. For the
sake of simplicity, only the case $E_f>E_i$ is schematically
represented in Fig.1. Two identical photons, of frequency $\omega
$, are absorbed in the process associated to the path labeled by
(a): the projectile gains an energy equal to $2\omega $, since the
internal state of the atom is not modified due to the scattering.
On the other path, (b), the high frequency photon is absorbed and
the fundamental one is emitted, leading to the same final energy:
$E_f=E_i+2\omega$. The relative phase $\varphi $ ''modulates'' the
quantum interference between the two paths, as it will be
discussed in this paper.

In the domain of high scattering energies and low field
intensities, we use the third order perturbation theory (second
order in the electric field and first order in the scattering
potential) to evaluate the differential cross section of the
scattered electrons. The calculations are described in the Section
II; they are carried out taking into account {\it all} the
involved Feynman diagrams for each of the paths (a) and (b). This
represents an appropriate treatment of two-photon free-free
transitions, including the modification of the target in the field
in second order perturbation theory. The case we study here
implies the quantum interference of two processes involving the
same numbers of photons, two.
{\it
In the energy spectrum of the scattered electrons this corresponds to the
second pair of sidebands ($N= \pm2$).}
It is the simplest case of this type in
which phase effects may be investigated taking consistently into
account the dressing of the target.
{\it Although it involves second order processes,
this case is more suitable for a discussion of interferences and phase
effects than the first pair of sidebands ($N=\pm1$).
In that case, the interferences would involve one- and
three-photon processes.}
Our numerical results,
presented in Section III, are obtained for identical linear
polarizations, in the geometry in which the polarization vector,
$\vec{\varepsilon}$, is parallel to the initial momentum of the
projectile. We consider fast projectiles, $E_i=100$ eV, and the
frequency $\omega =1.17$ eV, corresponding to Nd:YAG laser. Phase
effects are investigated in the domain of small scattering angles,
pointing out the influence of target dressing on these effects.
\nopagebreak[3]
\section{Basic Equations}

The time evolution of the electron-hydrogen system in the presence of the
electromagnetic field described by Eq.(\ref{c2}) is governed by the
hamiltonian
\begin{equation}
{\cal H} =  \frac {{\vec P}^2}{2} -\frac{1}{R} + \frac{{\vec
p}^{\;2}} 2 +
\frac{1} {\vert\vec{r} - \vec{R}\vert} - \frac{1}{r} + \frac 1 c \left[ \vec{%
p} + \vec{P} \right] \cdot \vec{{\cal A}}(t)
\equiv  H_0 + V + W(t) ,  \label{ha}
\end{equation}
where $\vec{R}$, $\vec{P}$ are the position and momentum operator of the
bound (atomic) electron and $\vec{r}$, $\vec{p}$ are the position and
momentum operator of the free (projectile) electron. $V \equiv -r^{-1} +
\vert\vec{r} - \vec{R}\vert^{-1}$ denotes the e-H interaction in the direct
channel, when exchange effects are neglected. $W(t) \equiv c^{-1} \left[
\vec{p} + \vec{P} \right] \cdot \vec{{\cal A}}(t)$ denotes the interaction
of the charge particles with the field, treated in the velocity gauge, using
the dipole approximation. The $\vec {{\cal A}}^2$-term was eliminated
through a unitary transformation.

In the {\it first nonvanishing order} of the perturbation theory,
the $S-$ matrix elements corresponding to two-photon processes are
given in second order perturbation theory by
\begin{equation}
S^{(2)} = - \int_{-\infty}^{+\infty}dt_1\int_{-\infty}^{t_1}dt_2 <{\chi}_f^-
|\tilde W(t_1) \tilde W(t_2) | {\chi}_i^+ > ,  \label{smat}
\end{equation}
where $\tilde W(t) = e^{iH_0t}W(t)e^{-iH_0t}$. In the previous
equation $ |\chi _i^{+}>$ and $|\chi _f^{-}>$ describe the initial
and final states of the colliding system (electron-atom)
\begin{eqnarray}
|\chi _i^{+}>& = &|\Psi _i>+G^{+}({\cal {E}}_i)V|\Psi _i>, \\
|\chi _f^{-}>& = &|\Psi _f>+G^{-}({\cal {E}}_f)V|\Psi _f>,
\end{eqnarray}
where
\begin{equation}
G^{\pm} ({\cal E}) = \left[{\cal E}-H_0-V\pm i\delta \right]^{-1}
\end{equation}
and $\delta$ a positive infinitesimal number. $|\Psi _{i,f}>$ are the
asymptotic states corresponding to the colliding system in the absence of
the interaction~$V$
\begin{eqnarray}
|\Psi _i>&=&|\psi _{1s}>|K_i>, \\
|\Psi _f>&=&|\psi _{1s}>|K_f>.
\end{eqnarray}
Here $|\psi_{1s}> $ denotes the ground state of a hydrogen atom
and $|K_{i,f}>$ are plane waves. The initial and final energies of the
electron-atom system are
\begin{eqnarray}
{\cal {E}}_i&=&{\rm E}_{1s}+\frac{p_i^2}{2}, \\ {\cal
{E}}_f&=&{\rm E}_{1s}+\frac{p_f^2}{2} ,
\end{eqnarray}
where ${\rm E}_{1s}$ is the unperturbed ground state energy and
$p_{i(f)}$ is the initial (final) momentum of the projectile.

Our goal is to study two-photon processes leading to the same
final energy of the scattered electron, given by equation
(\ref{ener}).
These processes are described by the following transition matrix element
\begin{eqnarray}
T_{if}^{(\pm 2)} &=&\frac{{\cal A}_0^2}4<{\chi }_f^{-}|{\vec{\varepsilon}}%
\cdot (\vec{p}+\vec{P})\;G^{+}({\cal {E}}_i\pm \omega )\;{\vec{\varepsilon}}%
\cdot (\vec{p}+\vec{P})|{\chi }_i^{+}>  \nonumber \\
&+&e^{\mp i\varphi }\;\frac{{\cal A}_0{\cal A}_0^{\prime }}4[<{\chi }_f^{-}|%
\vec{\varepsilon}\cdot (\vec{p}+\vec{P})\;G^{+}({\cal {E}}_i\pm 3\omega )\;{%
\vec{\varepsilon}}^{\;\prime }\cdot (\vec{p}+\vec{P})|{\chi }_i^{+}>
\nonumber \\
&&\hspace*{1.5cm}+<{\chi }_f^{-}|{\vec{\varepsilon}}^{\;\prime }\cdot (\vec{%
p}+\vec{P})\;G^{+}({\cal {E}}_i\mp \omega )\;\vec{\varepsilon}\cdot (\vec{p}+%
\vec{P})|{\chi }_i^{+}>],  \label{tm}
\end{eqnarray}
which is related to the $S$-matrix element (\ref{smat}). The upper
sign corresponds to the process in which the energy of the
scattered electron is increased by 2$\omega $ and the lower sign
to the case in which the energy is decreased by 2$\omega $. For
the sake of simplicity, we discuss here the significance of the
two terms in Eq.(\ref{tm}) only for the case $E_f>E_i$,
represented in Fig.1. The first line in equation (\ref {tm}),
which is proportional to the intensity of the fundamental field,
represents the transition matrix element corresponding to the
process involving the absorption of two {\it identical } photons
and we denote it by $T_a$. The other term is proportional to the
potential vector of both components of the field (\ref{c2}) and it
is connected to the diagrams in Fig.1(b). It involves two {\it
different} photons: the harmonic photon is absorbed and the
fundamental one is emitted; we denote it by $T_b$.
{\it Note that any high order correction to these leading terms are,
at least, of the fourth order in the field.}
A similar
analysis may be done for $E_f<E_i$. In order to describe the
process we are interested in, we match the individual matrix
elements, $T_a$ and $T_b$; the match involves the {\it relative
phase} $\varphi $. One may write
\begin{equation}
T_{i,f}^{(\pm 2)}=T_a+e^{\mp i\varphi }T_b.  \label{match}
\end{equation}

The evaluation of individual matrix elements was already
extensively discussed in the literature; it is based on the
'two-potential' formalism used by Kracke {\it et al} (1994) for
two identical photons and by Cionga and Buic\u{a} (1998) for two
different photons. In these works the authors restrict themselves
to the domain of high scattering energies, therefore the first
Born approximation is used to treat electron-atom scattering. In
this way, the evaluation of the transition matrix element is made
in the third order perturbation theory: the second order in the
electric field and the first order in the scattering potential,
$V$. When all the involved Feynman diagrams are included, every
transition matrix element for a two-photon process may be written
as the sum of three terms. For example, the process given in
Fig.1(a) is described by
\begin{equation}
T_a=T_P^a+T_M^a+T_A^a  \label{tmat}
\end{equation}
and a similar relation may be written for the process in Fig.1(b). $%
T_P^{a(b)}$, $T_M^{a(b)}$, and $T_A^{a(b)}$ account for the
electronic, mixed, and atomic contributions, respectively. Each
contribution is connected to specific Feynman diagrams as
discussed by Kracke {\it et al} (1994) for identical photons and
by Cionga and Buic\u{a} (1998) for different photons. The angular
structure of the these contributions, as well as their dependence
on the frequencies and on the momentum transfer, are analyzed in
the same papers. We mention that the analytic expression of $T_a$
is the same for $\Delta E_{\pm}=E_f-E_i=\pm 2\omega $. However,
for the two different signs of $\Delta E_{\pm} $ this expression
is computed using different values of the parameter of the Green's
function in Eq.(\ref{tm}). The same statements are true for $T_b$.

The differential cross section for the scattered electrons with
the final energy
(\ref{ener}) 
is then given by
\begin{equation}
\frac{d\sigma (\pm 2)}{d\Omega }={(2\pi )}^4\frac{p_f}{p_i}{|T_{if}^{(\pm
2)}|}^2,  \label{sec}
\end{equation}
being proportional to
\begin{equation}
|T_{if}^{(\pm 2)}|^2=|T_a|^2+|T_b|^2+2Re\left( T_a^{*}T_be^{\mp i\varphi
}\right) .
\end{equation}
The third term in the last expression is the interference one and
it depends on the phase difference, $\varphi $. For bichromatic
fields whose frequencies satisfy the relation $ 2\omega <|E_{1s}|$
both individual matrix elements, $T_a$ and $T_b$, are real (Cionga
and Buic\u{a} 1998), therefore the phase dependence is simpler:
\begin{equation}
\frac 1{I^2}\frac{d{\sigma }(\pm 2)}{d\Omega }={(2\pi )}^4\frac{p_f}{p_i}%
\left[ {\cal T}_a^2+\frac{I^{\prime }}I{\cal T}_b^2+2\sqrt{\frac{I^{\prime }}%
I}{\cal T}_a{\cal T}_b\cos {\varphi }\right] .  \label{sec.bi}
\end{equation}
In this relation we have chosen to explicitly display the
intensity dependence of the differential cross section, using the
relations $T_a=I\; {\cal T}_a$ and $T_b=\sqrt{II^{\prime }}{\cal
T}_b$. $I$ is the intensity of the laser field and $I^{\prime }$
that of the harmonic. Due to the power low, valid in the
perturbative regime, we prefer to normalize our results to the
square of the laser intensity. We note that in this case the
differential cross section (\ref{sec.bi}) is symmetric with
respect to $\varphi =\pi $.

\section{Results and Discussion}

In order to investigate the effects of the relative phase on laser
assisted signals in free-free transitions, we carried out
numerical calculations of the differential cross section
(\ref{sec.bi}) for fast projectiles, $E_i$ = 100 eV, in the domain
of optical frequencies. We illustrate our results for the
frequency of Nd:YAG laser, $\omega $ = 1.17~eV. Both components of
the field (\ref{c2}) are linearly polarized, namely
${\vec{\varepsilon}}={\vec{\varepsilon}} ^{\;\prime }\equiv
{\vec{p}}_i/p_i$;  ${\vec{\varepsilon}}$ defines the $Oz$ axis. We
have focused our attention on the study of phase effects at small
scattering angles, where the dressing of the target is important
and all three terms
in Eq.(\ref{tmat}) do contribute (Kracke {\it et al} 1994, Cionga
and  Buic\u{a} 1998). The numerical evaluation of the individual
transition matrix elements $T_a$ and $T_b$ is based on
analytic expressions involving series of hypergeometric functions
(Cionga and Florescu 1992, Cionga and Buic{\u{a}} 1998).

Figures 2(a) and (b) are three dimension plots:
the differential cross sections
(\ref {sec.bi}) are shown as a function
of the scattering angle, $\theta $, and of the relative phase,
$\varphi $, for equal intensities, $I=I^{\prime }$.
{\it The panel (a) corresponds to $\Delta E_+=2\omega$ and the
panel (b) to $\Delta E_-=-2\omega$.}
In order to
understand the ''modulations'' of these ''surfaces'', we note that
the general structure of the differential cross section
(\ref{sec.bi}) is given by
\begin{equation}
\frac{1}{I^2}\frac{d{\sigma (\pm 2)}^{}}{d\Omega }\sim {\cal
L}(\theta )+{\cal L}^{\prime }(\theta )\cos \varphi ,  \label{ll}
\end{equation}
where
\begin{eqnarray}
{\cal L}(\theta ) & = &{\cal T}_a^2+\frac{I^{\prime }}I{\cal
T}_b^2, \nonumber \\
{\cal L}^{\prime }(\theta ) & = &2\sqrt{\frac{I^{\prime }}I}{\cal T}_a{\cal %
T}_b,
\end{eqnarray}
with ${\cal L}(\theta )\geq 0$.  As well as the individual matrix
elements, $T_a$ and $T_b$, these two quantities
depend on the scattering angle. They also depend on the momentum
transfer of the projectile and on the field frequencies, $\omega $
and $3\omega $.
{\it The two deep minima present in Figs.2 (a) and (b)
for the same relative phase, $\varphi=180^0$, occur because
${\cal L}\simeq {\cal L}^{\prime }$ for $\theta \simeq 11^0$
when $\Delta E_+=2\omega$ and for $\theta \simeq 6^0$
when $\Delta E_-=-2\omega$.}

More information is revealed in figure 3, which displays
the differential cross section,
$d\sigma(+2)/d\Omega/I^2$, as a function of the
relative phase, $\varphi $, for six values of the scattering
angle, belonging to the domain where the dressing is important.
The parameters are the same as in Fig.2(a). For a given scattering
angle, when the harmonic is out of phase with respect to the
fundamental ($\varphi \neq 0 $), the laser assisted signal is
increased or decreased depending on the sign of ${\cal L}^{\prime
}$, as shown by Eq.(\ref{ll}). For all $\theta $ values for which
one of the individual transition matrix elements, $T_a$ or $ T_b$,
vanishes due to a dynamical interference between the three terms
in Eq.(\ref{tmat}), the laser assisted signal is $\varphi
$-independent. This is the case for example at $\theta \simeq
13.4^{{\rm 0}}$, where ${T}_b=0$. The first three changes of
curvature between the panels (a) and (b), (b) and (c), and (c) and
(d) are due to the three zeroes of the other matrix element,
$T_a$; they are located at $\theta \simeq 4.4^0,6.4^0$ and
10.25$^0$, respectively. More general, the $\varphi $-dependence
change its curvature after every zero of the individual matrix
elements that is due to cancellation between electronic, mixed,
and atomic contributions (\ref{tmat}).
{\it There are only two changes of curvature in Fig.2(b)
because only two such cancellations take place if $\Delta E_-=-2\omega$,
one for $T_a$ and the other one for $T_b$.}

The features discussed in the previous paragraph are the signature
of the dressing of the target. When this dressing is neglected,
the shape of the curve giving the $\varphi $-dependence does not
depend on the scattering angle, as one can see following the
dotted curves in Fig.3. These curves represent
$d\sigma(+2)/d\Omega/I^2$
computed in the approximation that
only the first term is kept in Eq.(\ref{tmat}) and its equivalent
for $T_b$. We remind that this approximation is the generalization
of Bunkin and Fedorov formula (1965) for a bichromatic field. Such
calculations were carried out by Varr\'o and Ehlotzky (1993a) in a
different regime of frequencies and energies, therefore no
specific comparison is made with that work. When the dressing is
neglected, the $\theta $-dependence factorizes out as follows:
\begin{equation}
\frac{1}{I^2} \frac{d\sigma(\pm 2)}{d\Omega} \simeq
\left[f_{el}^{B1}(q)\right]^2
 \frac{|(\vec{\varepsilon}\cdot \vec{q})|^4}{2^6\omega^8}
 \left[ 1+\frac {4}{81} \frac{I^{\prime }}{I}
        -\frac{4}{9}\sqrt{\frac{I^{\prime }}I}\cos {\varphi }\right]
,\label{v-e}
\end{equation}
where $\vec{q}$ is the momentum transfer of the projectile and
$f_{el}^{B1}=2\left(q^2+8\right)/\left(q^2+4\right)^2$ is the first
Born approximation of the transition amplitude for elastic
scattering on the potential $V$. We point out that, for $\theta
=\arccos (p_i/p_f)$, the scalar product $\vec{\varepsilon}\cdot
\vec{q}$ vanishes and therefore the differential cross section
(\ref{v-e}) vanishes, too. At large scattering angles, where the
target dressing does not contribute significantly anymore, the full
and the dotted curves become closer and closer, as one can see for
$\theta$=20$^0$.

An important parameter that has an influence on the quantum
interference between the two paths in Fig.1 is the ratio between
the intensities of the two field components, as one can see from
Eq.(\ref{sec.bi}). In Figure 3, when the dressing is neglected and
$I=I^{\prime}$, the laser assisted signal (dotted curves) is
always increased if $\varphi \neq 0$. For other ratios $I^{\prime
}/I$ the $\varphi$-dependence may be more complicated. In Figure 4
we have represented $d\sigma(+2)/d\Omega$,
normalized with respect to $I^2$,
as a function of the scattering angle, $\theta $, for four
different values of the relative phase: $\varphi =0;\pi /4;\pi
/2;\pi $. In each panel there are three curves, which correspond
to the following intensity ratios: $I'/I$ = 1; 10$^{-1}$;
10$^{-2}$. When $\varphi =0$ and this ratio gets smaller and
smaller, the differential cross section becomes closer and closer
to that given by the path (a) in Fig.1. In particular, for $I'/I=
10^{-2}$ the three zeroes of $T_a$ discusses previously can be
seen at the locations mentioned before. On the contrary, with the
increasing of the harmonic intensity, one should recover the
differential cross sections given by the path (b).

\section{Conclusion}
In this paper we have studied the effect of the relative phase
between the harmonic ($3\omega$) and the fundamental field
($\omega $) on two-photon free-free transitions in laser-assisted
electron-hydrogen scattering. Using third order perturbation
theory and taking into account all the involved Feynman diagrams,
we have evaluated the differential cross sections for scattered
electrons of energy $E_f=E_i\pm2\omega$.
The interference between the two quantum paths
leading to the foregoing final energies was modulated by the relative
phase. The signature of the target modification in the bichromatic
field was discussed for
fast projectiles and low field intensities
in the domain of small scattering angles, where the dressing
is important.
We stress that whenever the
target dressing can not be neglected, it influences significantly
the phase effects.

\vspace*{.5cm}

\noindent Bunkin, V. \& Fedorov, M. V. 1965 Zh.Eksp.Teor.Fiz.49,
1215 [1966 Sov.Phys. JETP22, 884].
\\ Cionga, A. \& Buic\u{a}, G. 1998 Laser Phys.8, 164.
\\ Cionga, A. \& Florescu, V. 1992 Phys. Rev. A45, 5282.
\\ Cionga, A. \& Zloh, G. 1999 Laser Phys.9, 69.
\\ Dubois, A. et al 1986 Phys. Rev. A34, 1888.
\\ Ehlotzky, F. 1994 Nuovo Cimento D16, 453 .
\\ L'Huillier et al, A. 1992 Adv. Atom. Mol. Opt. Phys. Suppl.1, 139 (ed. M.
Gavrila).
\\ Kracke, G. et al 1994 J. Phys. B27, 3241.
\\ Varr{\'{o}}, S. \& Ehlotzky, F. 1993 Phys. Rev. A47, 715.
\\ Varr{\'{o}}, S. \& Ehlotzky, F. 1993a Opt. Commun. 99, 177.
\\ Varr{\'{o}}, S. \& Ehlotzky, F. 1997 J. Phys. B30, 1061.
\\ V{\'{e}}niard, V. et al 1996 Phys. Rev. A54, 721.


\newpage

{\bf Figure Captions}

\begin{itemize}
\item Fig.1: Energy diagrams schematically
representing two-photon free-free transitions between the initial
state, in which the projectile has the energy $E_i$, and the final
one, in which it has the energy $E_f=E_i+2\omega$. (a) corresponds
to the absorption of two photons of frequency $\omega $. (b)
corresponds to the absorption of the harmonic and the emission of
the laser photon. The laser photons are represented by thin lines,
the harmonic photons by thick lines.
\item Fig.2(a): $d\sigma(+2)/d\Omega/I^2$
in logarithmic scale, as a function of the
scattering angle, $\theta $, and the relative phase, $\varphi $.
The initial energy is $E_i=100$ eV and the laser frequency is
$\omega=1.17$ eV. The harmonic intensity is chosen equal to that
of the laser. (b), idem fig. 2(a) but $d\sigma(-2)/d\Omega/I^2$.
\item Fig.3:  $d\sigma(+2)/d\Omega/I^2$
in logarithmic scale, as a function of the
relative phase, $\varphi $, is represented by full lines for six
values of the scattering angle. The dotted lines represent the
same quantity when the target dressing is completely neglected.
The parameters are the same as in Fig.2.
\item Fig.4: $d\sigma(+2)/d\Omega/I^2$
in logarithmic scale, as a function of the
scattering angle, $\theta $, for four values of the relative phase:
$\varphi = 0$,  $\varphi =\pi /4$, $\varphi =\pi /2$, and $\varphi
=\pi $.  The initial energy is $E_i=100$ eV, and the laser
frequency is $\omega=1.17$ eV. The intensity of the harmonic field
represents 1\% of the laser intensity (full lines), 10\%
(dotted-dashed lines), and 100\% (dashed lines).
\end{itemize}

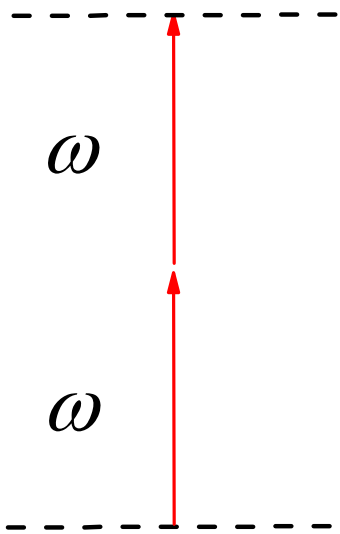
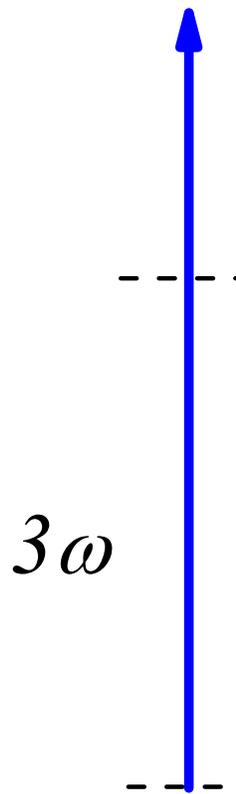
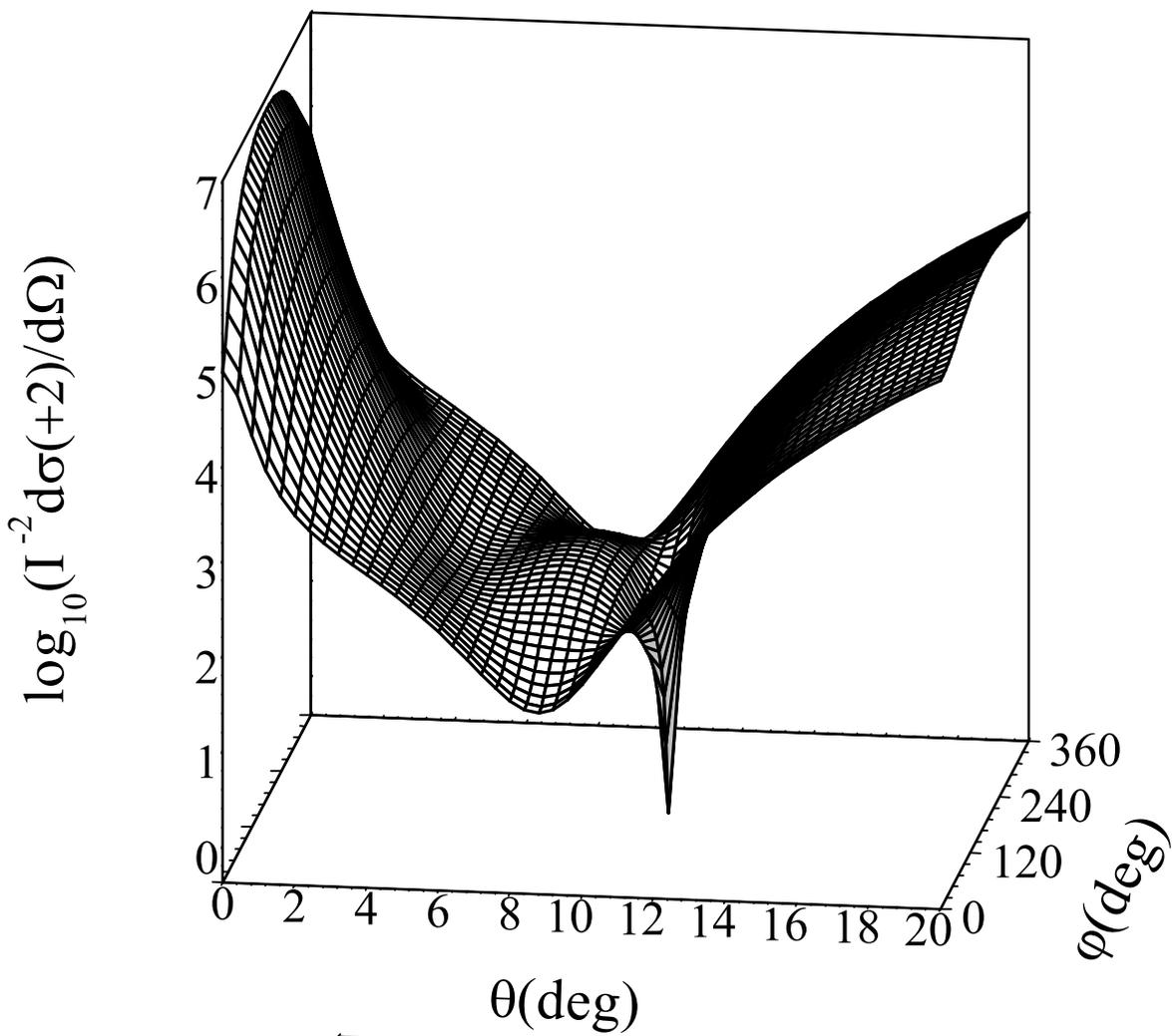
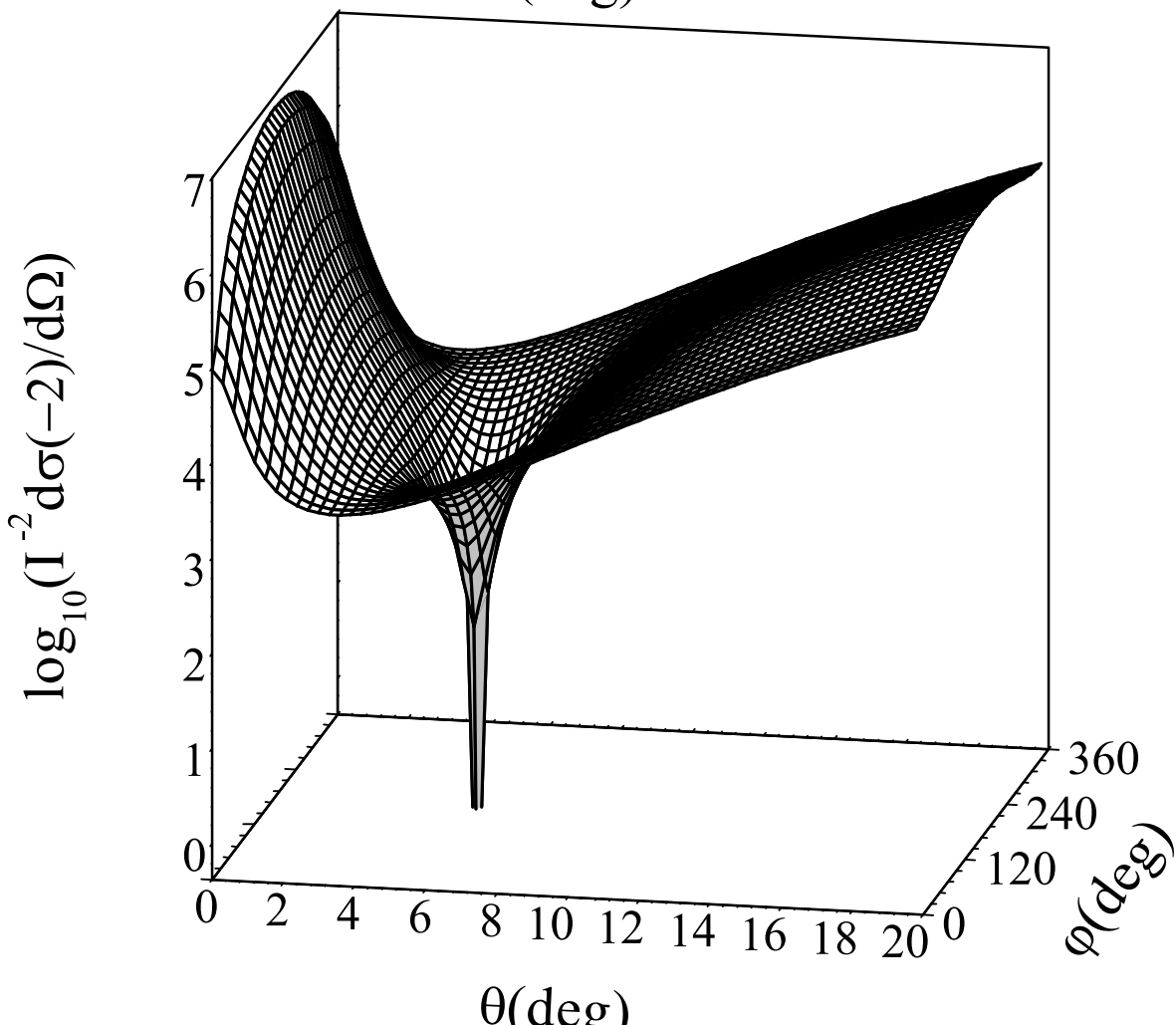
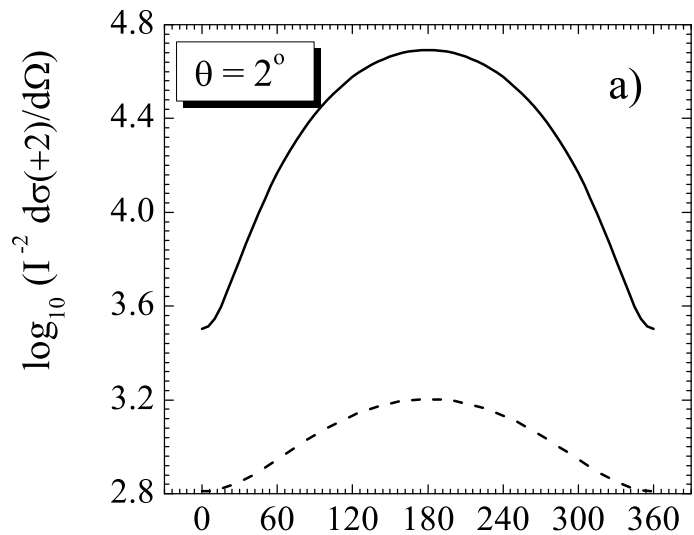
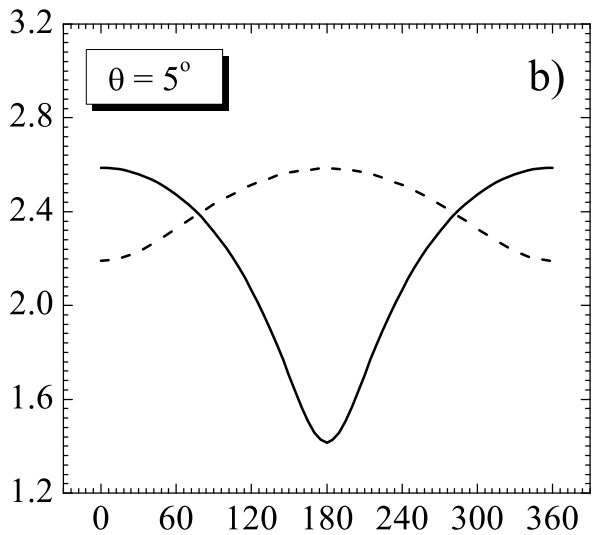
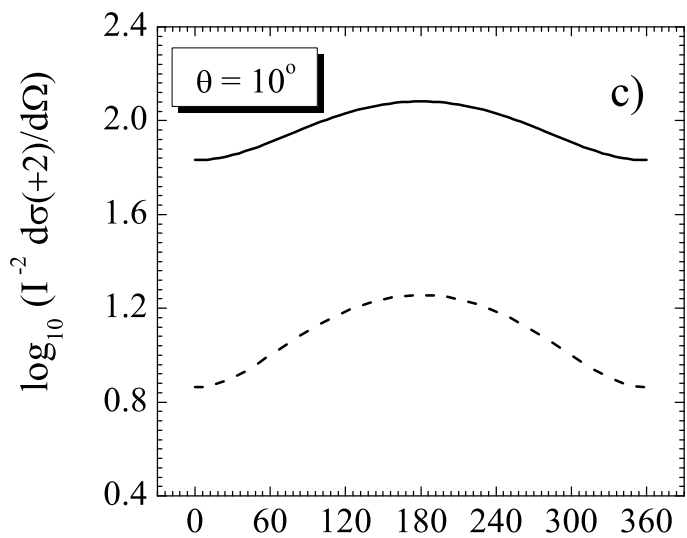
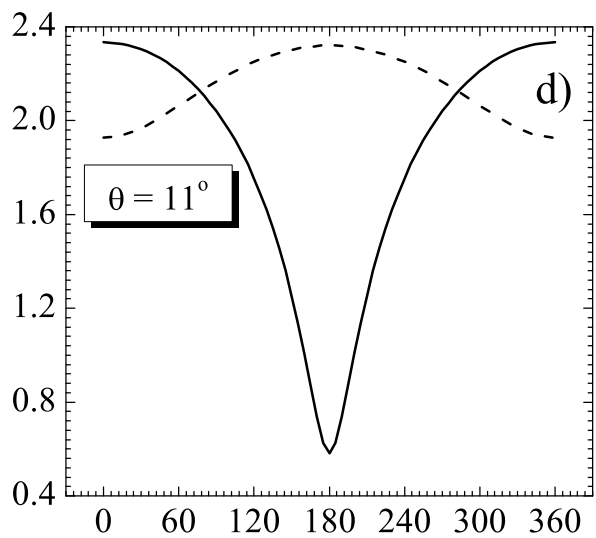
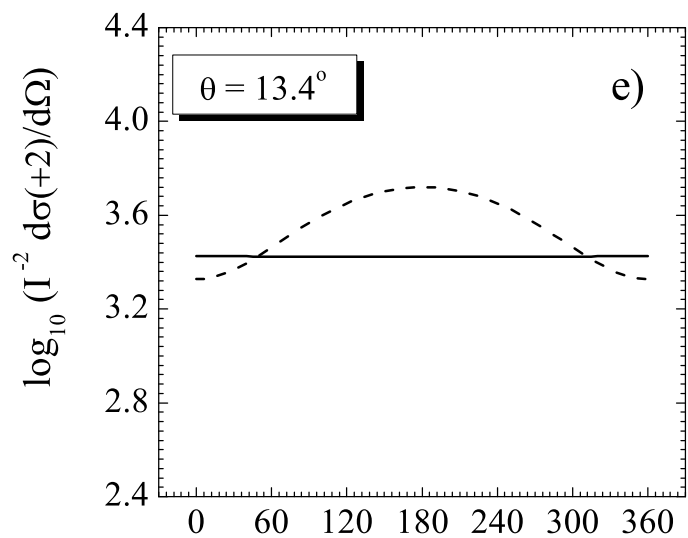
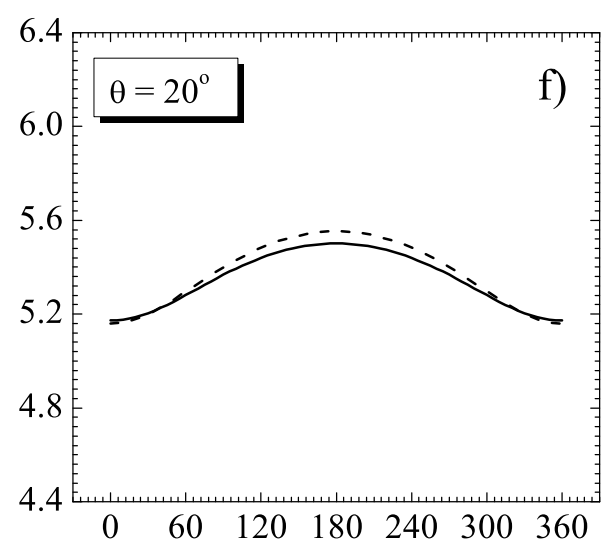
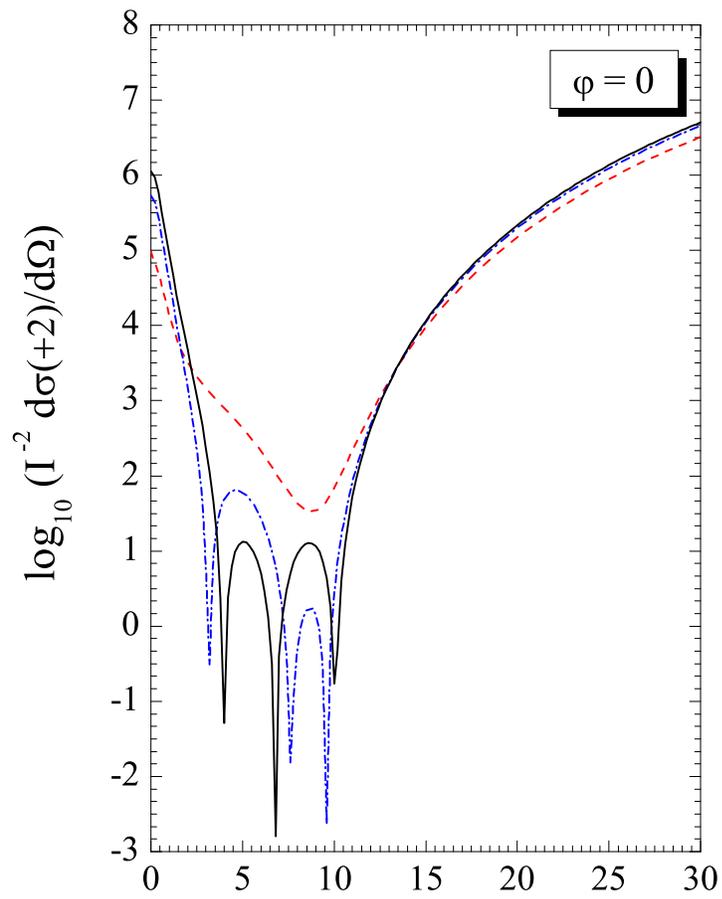
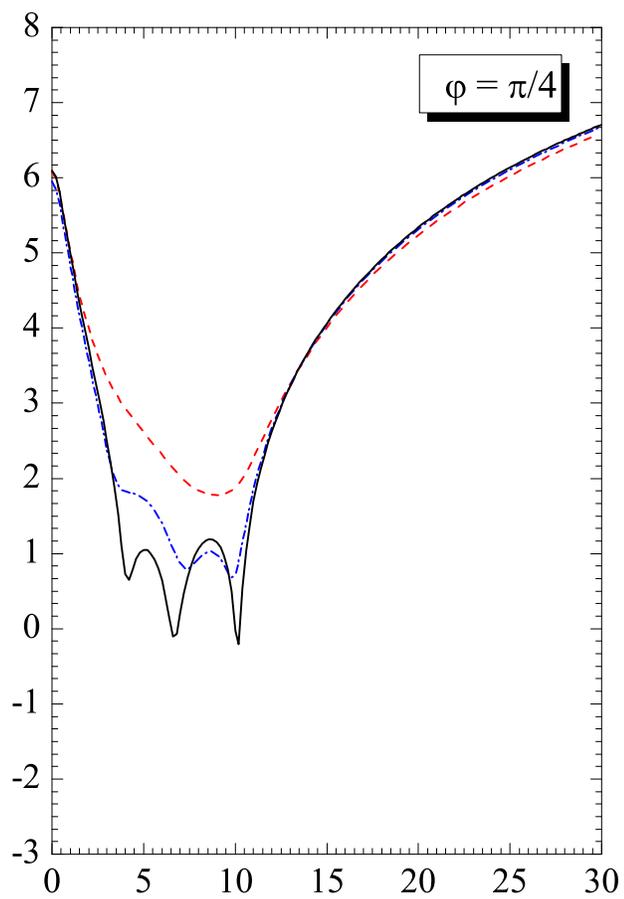
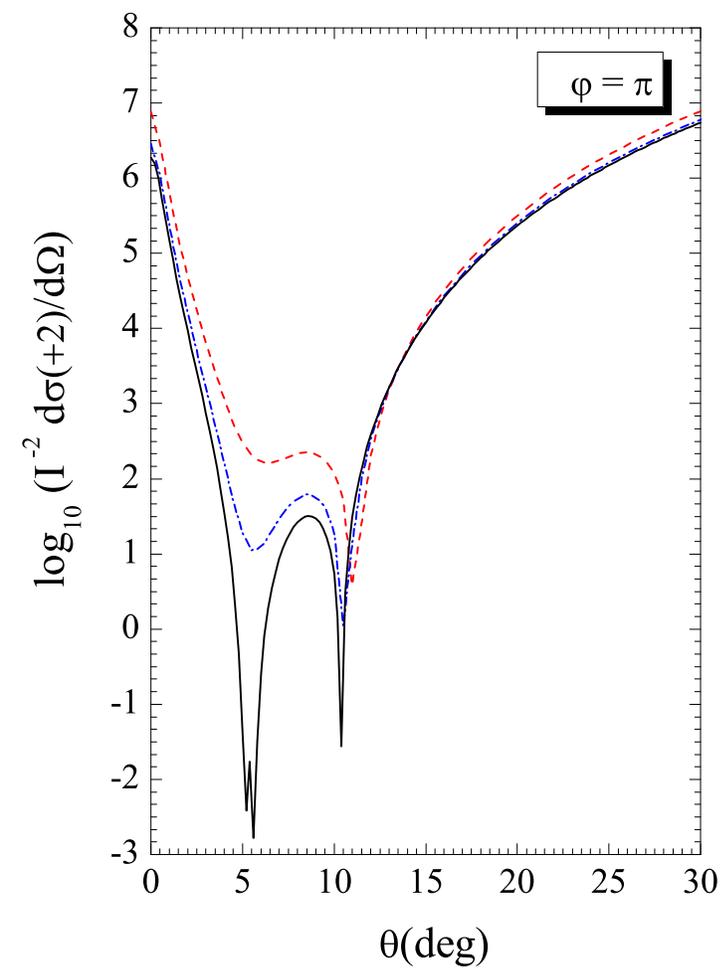
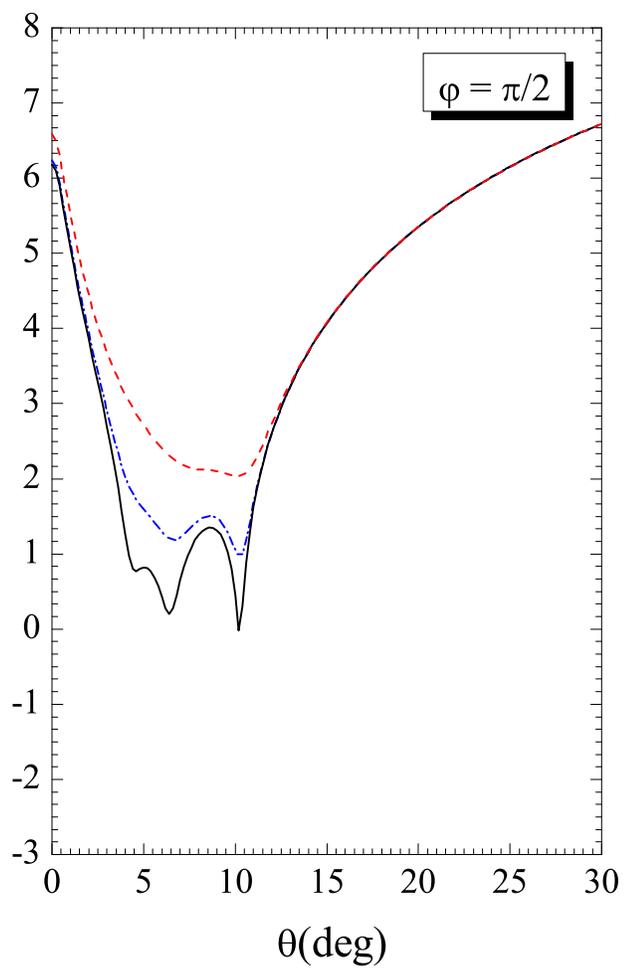
\end{document}